\documentclass{article}
\usepackage{graphicx} 
\usepackage[english]{babel}
\usepackage[letterpaper,top=2cm,bottom=2cm,left=3cm,right=3cm,marginparwidth=1.75cm]{geometry}

\usepackage{soul,color}
\usepackage{authblk}
\usepackage{amsmath}
\usepackage{amssymb}
\usepackage{amsfonts}
\usepackage{bm}
\usepackage{graphicx}
\usepackage{siunitx}
\usepackage{lineno}
\usepackage{caption}
\usepackage[colorlinks=true, allcolors=blue]{hyperref}

\title{Freeform Programmable Liquid Crystal Elastomers via Monolithic Integration of Co-Optimization and 4D-Printing}

\author[1,$\dagger$]{Yibo Wang}
\author[1,$\dagger$]{Asad Farooq Ahmed}
\author[1]{Weichen Li}
\author[1]{Weiyun Xu}
\author[1,2,3,*]{Xiaojia Shelly Zhang}

\affil[1]{Department of Civil and Environmental Engineering, University of Illinois Urbana-Champaign, Urbana, IL 61801, USA}

\affil[2]{Department of Mechanical Science and Engineering, University of Illinois Urbana-Champaign, Urbana, IL 61801, USA}

\affil[3]{National Center for Supercomputing Applications, Urbana, IL 61801, USA}

\affil[$\dagger$]{Equal contribution}

\affil[*]{Corresponding author: X.S. Zhang (zhangxs@illinois.edu)}

\date{}

\begin{document}

\maketitle
\begin{abstract}

\noindent
Liquid crystal elastomers (LCEs) are active materials capable of large, reversible thermally induced actuation, offering significant potential for applications such as artificial muscles and soft robotics. 
However, their broader deployment is hindered by the lack of a unified design-to-fabrication framework, the strong coupling among geometry, director fields, and manufacturing constraints, and the difficulty of reliably translating complex designs into physically realizable architectures.
Here, we present an autonomous and monolithic integration of co-optimization and 4D printing that enables the end-to-end design-to-realization of {versatile} programmable LCE architectures with  {superior} precision, complexity, and versatility. 
The proposed framework establishes 
a unified design strategy that co-optimizes the freeform multilayer geometry and fully continuous layer-specific director distributions, and subsequently translates the optimized fields into fabrication-ready DIW toolpaths through an automated manufacturing-aware pipeline, promoting DIW compatibility and targeted functionality in three dimensions. To bridge computation and realization, we develop an optimization-centered 4D printing approach with automated, fabrication-aware toolpath generation tailored for direct ink writing of LCEs.
With the proposed design-to-fabrication paradigm, we experimentally realize architectures that exhibit complex, non-intuitive thermo-mechanical behaviors, including contraction-induced tunable global area expansion, programmable morphing into 
prescribed complex three-dimensional surfaces, and precise trajectory control of compliant mechanisms. 
Experiments achieve highly challenging prescribed functionalities beyond the reach of intuition-based approaches and show excellent agreement with numerical simulations across all samples.
This work establishes a generalizable inverse design and physical realization of freeform programmable LCE systems, enabling function-oriented and optimization-driven intelligent LCE applications for next-generation soft robotic and responsive material engineering.

\end{abstract}

\vspace{10 pt}

\section*{INTRODUCTION}

Active materials such as hydrogels~\cite{peppasHydrogelsBiologyMedicine2006}, piezoelectric materials~\cite{chorsiPiezoelectricBiomaterialsSensors2019}, and liquid crystal elastomers (LCEs)~\cite{ulaLiquidCrystalElastomers2018, chenBioinspiredMechanochromicLiquid2024} can actuate in response to environmental stimuli such as change in humidity, temperature, and applied magnetic or electric fields.
As a thermally active soft material, LCEs can produce large and reversible thermally induced actuation strain, which has enabled a wide range of capabilities and applications, including programmable morphing~\cite{wangAutonomousCodesignFabrication2025,peng4DPrintingFreestanding2022}, artificial muscles~\cite{liuShapeprogrammableDeformationlockingSelfsensing2022}, and robotic actuators~\cite{hebnerLeapingLiquidCrystal2023}. 
The large thermal actuation of LCEs arises from the entropic elasticity of the lightly cross-linked mesogens \cite{warnerLiquidCrystalElastomers2007}, which, upon temperature rise, shifts from the nematic state with orientation order to the isotropic state where the order is lost \cite{warnerLiquidCrystalElastomers2007,warner1991elasticity}.
This transition manifests as a large contraction along the nematic director macroscopically.
Upon cooling, the material recovers its original shape, thereby the actuation is fully reversible.

The key to achieving various thermal actuation of an LCE structure lies in the strategic design of the director distribution and structural geometry~\cite{wareVoxelatedLiquidCrystal2015,guo2021shape}.
Through careful designs, advanced capabilities such as adaptive shape morphing~\cite{barnesDirectShapeProgramming2019, barnesReactive3DPrinting2020, kotikian3DPrintingLiquid2018, zuo2019visible} and robotic locomotion~\cite{maurinLiquidCrystalElastomer2024, song2025frequency, palagiStructuredLightEnables2016, chenLightdrivenDandelioninspiredMicrofliers2023, chenScalableJetSwimmer2025} have been successfully achieved. The design of most LCE structures has primarily been based on intuition and experience, often requiring trial-and-error to acquire the desired functionalities and may be challenging to achieve more sophisticated ones. Inverse LCE design is often driven by physics-based algorithms, enabling efficient function-oriented designs that can realize more complex responses, such as target shape fitting~\cite{aharoniUniversalInverseDesign2018, xiaInverseDesignAxisymmetric2021}, programmable shape transformations~\cite{wangAutonomousCodesignFabrication2025,kowalskiCurvatureDesignDemand2018}, and tunable nonlinear mechanical responses~\cite{chengInverseDesignSmart2025}.
However, inverse design methods may rely on trained parameter–property models (e.g., surrogate or machine learning models) tailored to specific functionalities, 
while the selected designable material parameters may constrain elemental-level design freedom.
Moreover, most existing inverse design approaches have not effectively and directly integrated LCE's fabricability, producing LCE designs that may be challenging to manufacture. These technical gaps have hampered the broader application of LCEs in many fields.

As an effective generative design approach, topology optimization (TO)~\cite{Bendsoe1988,bendseTopologyOptimization2007,wang2021comprehensive} features high design freedom and great compatibility with many physics phenomena. For example, TO has been widely used to program structures with nonlinear responses~\cite{liDigitalSynthesisFreeform2022,liProgrammingPhysicalRealization2023} and tunable mechanical properties~\cite{clausenTopologyOptimizedArchitectures2015a,wangDesignMaterialsPrescribed2014b}, while also incorporating multiphysics functionalities such as magnetic actuation~\cite{zhaoTopologyOptimizationHardmagnetic2022,wangInverseDesignMagnetoactive2023,perez2025topology}, piezoelectric coupling~\cite{zhangFlexoelectricNanostructureDesign2022,chenTopologyOptimizationFlexoelectric2021}, and thermal responsiveness~\cite{chungLevelsetTopologyOptimization2020,wangMultimaterialTopologyOptimization2023}.
Integrating TO with LCEs is a promising pathway toward realizing complex LCE behaviors. 
Most studies in TO-based design of LCEs often approximated the spontaneous deformation as thermal expansion
, generating deformable bending composites~\cite{athinarayanaraoComputationalDesign4D2023}, layered origami structures~\cite{fuchiTopologyOptimizationDesign2015}, and LCE networks~\cite{wangAutonomousCodesignFabrication2025}, while enabling functionalities such as target surface fitting~\cite{yanTopologyOptimizationAutomated2022a} and enhanced energy absorption~\cite{barreraLiquidCrystalOrientation2024b}.
To better capture the complex behaviors of LCE, such as semi-soft elasticity, Li et al.~\cite{liArbitraryCurvatureProgramming2023} integrated a statistical mechanics-based LCE model with multimaterial TO. Through a curvature-based inverse design formulation, they achieved shape programming and temperature-switchable functionalities~\cite{liArbitraryCurvatureProgramming2023,liComputationalMorphogenesisLiquid2024,wang2026topology} for 2D LCE composites with planar behaviors. 
{To capture the time-dependent behavior of LCEs, Xu et al.~\cite{xu2025computational} optimized director distributions in viscoelastic LCE structures to achieve desired mechanical performance.}
\textcolor{black}{Despite several technical progress, most studies on TO-based design of LCEs remain limited in both the design capability and the realization of practically deployable and experimentally demonstrated functional systems.
Moreover, existing studies primarily focus on tailoring LCE director distributions, while the coupled role of geometry in enabling more sophisticated and diverse functionalities remains largely unexplored.}

On the fabrication front, inversely designed LCEs require high-fidelity manufacturing for their irregular director distribution, the key to which is the accurate alignment of mesogens to the optimized director field. For simple patterns, such alignment can be realized through mechanical stretching ~\cite{kupfer1991nematic,camacho2004fast,yakacki2015tailorable}, magnetic field control ~\cite{buguin2006micro}, and varying the cross-link density~\cite{ware2015programmed}. More complex patterns and geometries usually require direct-ink writing (DIW) for accurate fabrication, where
partially cross-linked LCE oligomers (ink) are shear-aligned as they are extruded through the nozzle, and the resulting director orientation is subsequently locked in by UV cross-linking immediately following the deposition ~\cite{wang20203DLCEmanu,kotikian20183d,ambulo2017four,lopez20184d,mistry2021soft}. \textcolor{black}{DIW is a strong candidate for the accurate realization of TO-based LCE inverse design featuring freeform geometry and complex director distributions. However, this has not been achieved thus far due to the lack of a function-oriented and DIW-integrated design strategy for LCE programming, thereby limiting the reach for more sophisticated and practically realizable LCE functionalities.}

\textcolor{black}{ 
This study develops a monolithic integration of multilayer co-optimization and DIW that enables the end-to-end design-to-fabrication of versatile LCE architectures for accurately programming highly complex yet practically realizable behaviors. Although existing inverse-design and topology-optimization frameworks have successfully realized various functionalities, they have primarily focused on two-dimensional LCE designs, adopted simplified models, focused only on the director's field optimization, or lacked experimental realization, thereby restricting the accessible functional space. Here, we advance this direction by developing a functionality-oriented, optimization-driven inverse-design framework based on a rigorous statistical-mechanics-based three-dimensional LCE model under large deformation, enabling the programming of complex three-dimensional actuation modes for previously unexplored functionalities. We greatly expand the design space by a multilayer, co-optimization of both the freeform geometry and continuous layer-specific director fields of LCEs. The director orientations are regularized using a continuous filtering technique to improve their compatibility with DIW fabrication. Tailored to DIW, we propose a topology-aware path-generation technique with an advanced nonlinear solution scheme and automated domain-partitioning technique to automatically convert organic optimized LCE geometries and complex director fields into smooth, continuous, and well-planned printing toolpaths readily usable for DIW, while improving alignment accuracy and compatibility with rapidly varying director patterns.
Together, the monolithically integrated design-to-DIW strategy enables us to synthesize and precisely fabricate optimized LCE structures with organic geometries and intricate director distributions that deliver a wide range of challenging two- and three-dimensional thermomechanical functionalities. Using an in-house-built DIW fabrication technique, we experimentally demonstrate (1) metastructures with programmable thermal expansion against the intrinsic local contraction of LCEs, (2) on-demand morphing of LCE sheets into arbitrarily prescribed three-dimensional surfaces, and (3) precise trajectory control and actuation behaviors enabling multifunctional compliant mechanisms and robotic motions. 
These functions are realized through new objective formulations within a unified design framework. 
Across the demonstrated functions, the inverse-designed LCE architectures accurately reproduce the target behaviors in both simulations and experiments. 
Beyond LCEs, the proposed functionality-driven co-optimization framework could be adapted to other anisotropic and stimuli-responsive material systems, including fiber-reinforced composites and magnetoactive materials.
The monolithically integrated design-to-fabrication workflow could likewise be extended to other extrusion-based additive-manufacturing processes requiring local directional control, such as fused deposition modeling (FDM), in which toolpath orientation governs local filament alignment and the resulting material anisotropy. 
More broadly, by integrating co-optimization with manufacturing-aware toolpath generation and DIW fabrication, the proposed strategy provides a general route for translating complex programmed functions into physical material architectures, with potential applications in soft robotics, wearable electronics, and adaptive systems.
}

\section*{RESULTS}

\subsection*{Digital-to-physical monolithically integrated co-optimization and DIW framework for LCEs}

\label{Subsec:TopDesign&Fab}
The proposed co-optimization inverse design strategy features rigorous physics models, freeform multilayer geometry, and continuous LCE director distribution, as illustrated in Figure~1a.
Specifically, the director orientations and freeform geometric features of the thermally-actuated LCEs are co-optimized with a multilayer configuration, where their director distributions are optimized independently while sharing the same geometry. To accurately capture the nonlinear thermo-mechanical large deformation of LCE, we employ a non-ideal LCE model
~\cite{biggins2008semisoft} and use high-fidelity nonlinear finite element analysis (FEA) for the numerical simulation of LCE's response (see Section 1 of the SI for details). 

To fully harness the design freedom, our autonomous design framework simultaneously optimizes the LCE structure's geometry (through density design variable $\boldsymbol{\rho}$) and the fully continuous director orientation (through design variable $\boldsymbol{\xi}$). 
\textcolor{black}{A density convolution filter followed by a Heaviside projection~\cite{wangProjectionMethodsConvergence2011} is applied to regularize the material distribution and promote well-defined geometries, following standard topology-optimization practice. 
To obtain smooth director distributions suitable for continuous DIW toolpath generation, we develop a continuous director-field filtering scheme to regularize spatial variations in orientation ({for details, see SI, Subsection 2.2}).
}
With this setup, the inverse-design problem is formulated as a target-matching optimization nested with nonlinear LCE equilibrium:
\begin{equation*}
\begin{aligned}
\min_{\boldsymbol{\rho},\,\boldsymbol{\xi}} \quad 
& \max_{k} \; 
\mathcal{E}_{k}\!\left(
\boldsymbol{u}_{k},
\boldsymbol{u}^{\mathrm{tar}}_{k}
\right) \\
\text{s.t.} \quad 
& \boldsymbol{r}_{k}(\boldsymbol{u}_{k},\boldsymbol{\rho},\boldsymbol{\xi})=\boldsymbol{0},
\quad k=1,\dots,N_{\mathrm{tar}}, \\
& \boldsymbol{g}(\boldsymbol{\rho},\boldsymbol{\xi}) \leq \boldsymbol{0}.
\end{aligned}
\end{equation*}
Here, $\boldsymbol{\rho}$ and $\boldsymbol{\xi}$ are the density and director design variables. 
The objective minimizes the maximum mismatch $\mathcal{E}_{k}$ between simulated and prescribed responses over $N_{\mathrm{tar}}$ target states, while $\boldsymbol{r}_{k}=\boldsymbol{0}$ enforces the nonlinear equilibrium of the LCE system at each state. 
The vector $\boldsymbol{g}$ collects case-specific design constraints, as detailed in Section~3 of the SI.
The optimization problem is solved with a gradient-based Method of Moving Asymptotes~\cite{svanberg1987mmamethod}, which iteratively updates the design toward the converged solution that achieves the prescribed target behaviors, as illustrated in Figure~1a.

\textcolor{black}{To smoothly convert the optimized LCE designs with organic geometries and complex director distribution into DIW-printable toolpaths, we develop a TO-aware path generation technology based on iso-contours of a scalar field with gradients perpendicular to the optimized director field.
The scalar field is obtained by solving a penalized Poisson equation via FEM (details in the SI, Section 4), from which iso-lines are extracted and trimmed to be aligned with the optimized geometry, as illustrated in the lower right on Figure~1a. Layer-wise implementation of this method based on continuous director patterns produces stacked DIW toolpaths designed to minimize gaps and excessive overlap.
For more sophisticated designs with rapidly varying director fields, we develop a new automatic domain-partitioning strategy for post-processing, which divides the domain into sub-regions of smoother director distributions, within which toolpaths are generated and subsequently merged.}
This seamlessly integrated design-to-fabrication workflow automatically preserves both freeform geometric and director information of optimized LCE designs while allowing for practical and smooth DIW manufacturing, facilitating reliable fabrication of complex optimized LCE samples required to achieve challenging functionalities.

For DIW of programmed LCE, as shown in lower left of Figure 1a 
, we synthesize LCE ink with the mixture of reactive monomers (RM-257), photo-initiator (I2959), chain extender (EDDET), and DCM (CH$_2$Cl$_2$).
During fabrication, LCE ink is loaded into a syringe on the DIW printer and printed at a speed of 120 mm/min (see Experimental  Section for details on fabrication). The LCE is printed at the nematic phase with a nozzle temperature of {$45\,^\circ\mathrm{C}$}.
The viscous shearing of the extruding ink aligns the director in the print direction at deposition, which is then immediately UV cross-linked. 
\textcolor{black}{The spontaneous thermal-actuation measurements shown in the left panel of Figure~2d indicate a nematic--isotropic transition temperature $T_{NI}$ of approximately $90\,^\circ\mathrm{C}$. 
When heated from room temperature to $140\,^\circ\mathrm{C}$, the uniformly aligned LCE strip specimens contract by $44.5 \pm 4.5\%$ along the printing direction, which coincides with the director orientation, and fully recover their undeformed configurations upon cooling, demonstrating reversible actuation. 
The right panel of Figure~2d presents the tensile stress--strain responses of LCE specimens with director orientations of $0^\circ$ and $90^\circ$ under vertical uniaxial tension. 
The stress plateau of the $90^\circ$ specimen reflects semi-soft elasticity associated with director rotation and may additionally contain viscoelastic contributions, highlighting the coupled reorientation of the mesogens and deformation of the polymer network.}

\begin{figure}
    \centering
    \includegraphics[width=0.7\textwidth]{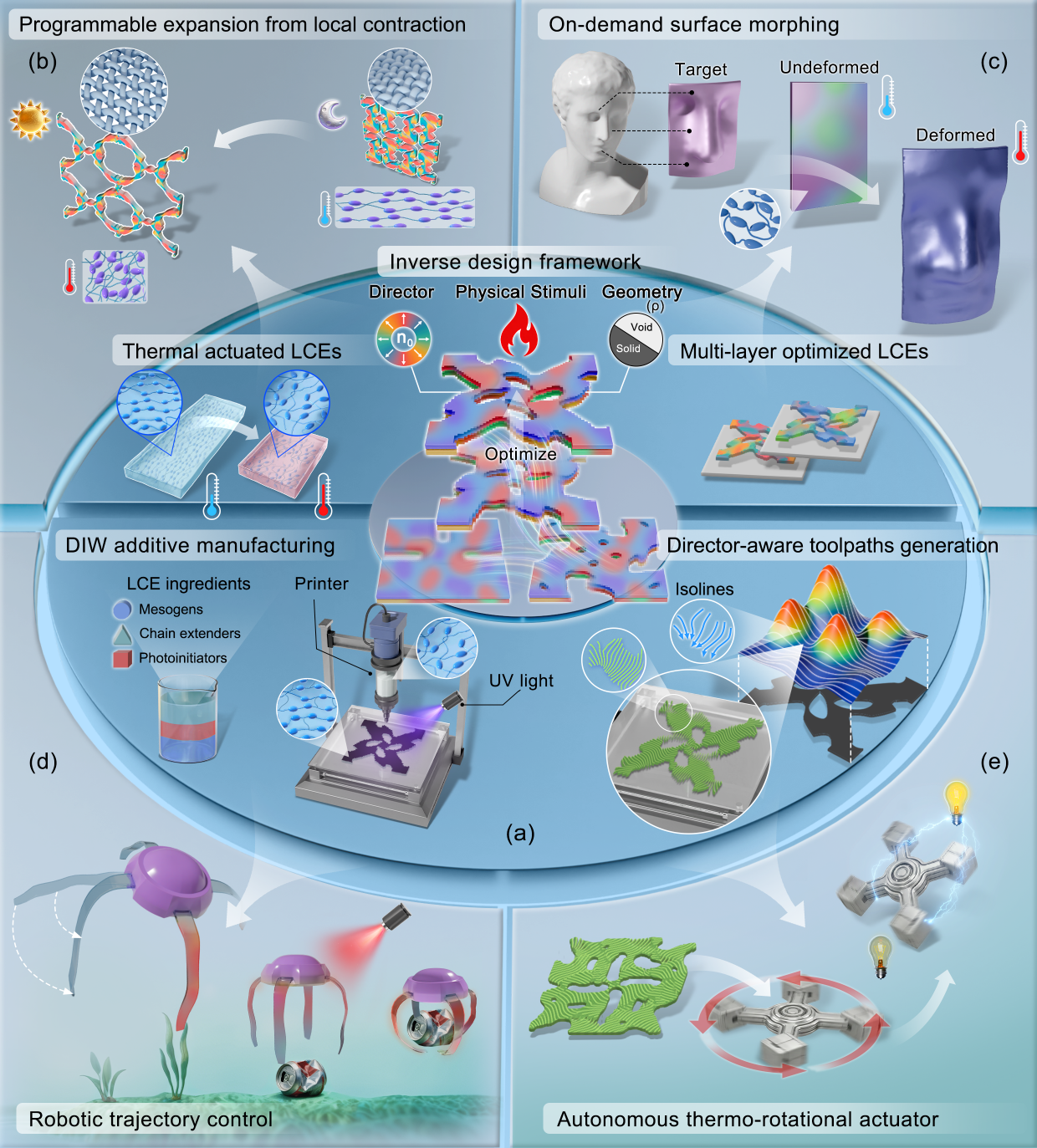}
    \caption*{%
    \textbf{\textcolor{black}{Figure~1. Autonomous LCEs inverse design-to-fabrication framework enabling versatile functionalities.}}
    (a)~ 
    Different sections jointly constitute the autonomous inverse design–fabrication framework for thermally responsive LCEs: entropy-driven order loss enables the nematic–isotropic transition and actuates LCE deformation upon heating; director orientation and geometry density are included as design variables for LCE architectures; LCEs are optimized to exhibit multilayer, distinct director patterns; DIW toolpaths are extracted from iso-contours of a regulated intermediate field to retain the desired director orientation within optimized geometries; physical LCE samples are fabricated by DIW using LCE ink.
    (b)~Functional LCE metastructures achieving Programmable thermal expansion from intrinsically contracting LCEs. 
    (c)~On-demand surface morphing of LCE architectures into arbitrary-form complex surface. 
    (d)~Octopus-inspired LCE soft robots enabling multifunctional actuated behaviors.
    (e)~Autonomous thermo-rotational LCE actuator realizing accurate prescribed trajectory.
    }
    \label{fig:Idea_overarching}
\end{figure}

\textcolor{black}
{
Methodologically, our framework moves beyond prior work on topology optimization of LCE by developing a multi-layer, simultaneous optimization of both structural geometry and layer-specific mesogen director fields, substantially expanding the available design space. The multilayer optimization formulation is further coupled with generalization to fully three-dimensional deformation, including complex out-of-plane morphing, as well as a continuous filtering technique that regularizes layer-specific director fields and promotes compatibility with layer-wise DIW fabrication. Moreover, inspired by field-based approaches~\cite{fang2024exceptional,chen2022field}, we introduced a new toolpath generation approach with a refined nonlinear formulation and automated post-processing strategies, enabling automatic conversion of optimized LCE geometries with complex director fields into smooth, continuous, and well-planned printing toolpaths for DIW. Various formulations for distinct functional objectives within a unified inverse-design framework are presented to demonstrate the generality of the approach. Collectively, these advances enable the DIW fabrication and physical realization of functional LCE architectures with optimized geometries and  complex director fields, which is rarely demonstrated in LCE design optimization studies.
} With the proposed strategy, we numerically and experimentally realize programmable expansive meta-structures driven by local mesogen contraction (Figure~1b), precise LCE shape morphing into an arbitrary, highly characteristic target surface in three dimensions (Figure~1c), creation of a bio-inspired amphibious soft robot with LCE arms to achieve multifunctional, regulated deformation through precise trajectory control (Figure~1d), and autonomous LCE rotator that functions as a thermal circuit for soft electronics (Figure~1e). 

\subsection*{Programmable thermal expansion metamaterial from inherently contracting LCEs}
\label{subsec:areaexpansion}

Accurate programming of area change (expansion and contraction) under non-mechanical stimuli is useful for applications ranging from intelligent fabrics (Figure~2a) to deployable 
thermal-shielding structures, to which LCEs hold great potential.
However, LCEs naturally undergo local contraction upon heating. Our topology optimization framework strategically harnesses structural geometry and director distribution to convert these intrinsic contractions into global expansion.
\textcolor{black}{To this end, we consider the setup of a periodic LCE unit cell structure, and formulate the objective function as minimizing the difference between the actual thermal expansion ratio predicted by FEA and the prescribed target: $\min_{\mathbf{\rho},\;\mathbf{\xi}} \tfrac{1}{2}\,\Bigl[J_{\mathrm{ave}}\bigl(\widetilde{\mathbf{u}}(\mathbf{\rho},\mathbf{\xi})\bigr) - J_{\mathrm{ave}}^*\Bigr]^{2}$, where $\widetilde{\mathbf{u}}$ denotes the planar displacement of LCEs under periodic boundary conditions and the plane-stress assumption, $J_{\mathrm{ave}}$ is the determinant derived from displacement quantifying the occupied area, and $J_{\mathrm{ave}}^*$ is the target determinant corresponding to the desired expansion ratio of a single LCE unit cell}, as indicated in Figure~2b.
Both material density $\mathbf{\rho}$ and director orientation $\mathbf{\xi}$ are optimized to achieve the specified area expansion, subjected to volume and stiffness constraints that prevent excessive material usage and maintain structural connectivity in the unit cell. The optimized unit cell is then periodically assembled into a larger LCE metamaterial that features the same thermal expansion.

With the specifically tailored optimization setup for LCEs, the on-demand programmability is validated via two distinct application scenarios that require different reaction forces.
\textcolor{black}{
Both design cases are initialized with a uniform density of $\rho=0.5$. Director orientations of $\pi/4$ are prescribed within the first quadrant of the square design domain and mapped to the entire domain through a centrosymmetric filtering operation.
}
In the first scenario, we specified a smaller target expansion ratio under a reduced stiffness constraint, suitable for applications requiring limited reaction forces and constrained expansion, such as medical implants~\cite{versteegden2018self}.  
The resulting two-dimensional (2D) expansive design of LCE (denoted as Dsg. 1) exhibits a cross-like braided structure, where diagonally interlinked lattices form the primary stretching arms with smoothly varying director distributions.
\textcolor{black}{Upon heating, local contraction along directors oriented perpendicular to the stretching-arm axes induces transverse extension of the arms along the diagonals, which lead to the expansion of the overall structure and yielding a numerical expansion ratio of 1.401.}
In the second scenario, we required a larger overall expansion subject to higher stiffness (i.e., greater reaction forces), corresponding to potential applications in heat-proof deployable structures~\cite{hu2025reconfigurable}.  
\textcolor{black}{
As illustrated in Figure~2c, Dsg.~2 evolves from the same uniform initial configuration. During optimization, the material progressively concentrates along the diagonals, while internal voids expand to form an interconnected lattice-like topology. After 95 iterations, the topology converges to a bio-inspired configuration resembling frog legs connected by several muscle-like segments. These segments predominantly contain continuous director fields oriented perpendicular to the muscle alignment. 
Upon thermal activation, contraction along the directors generates transverse extension along the axes of the muscle-like segments, progressively unfolding the limb-like structure to expand the architecture and yielding a numerical expansion ratio of 1.534.
In contrast to uniformly aligned LCE strips, which contract along their directors upon heating (Figure~2d), the optimized architecture combines an intricate geometry with spatially programmed director fields to generate a bio-inspired frog-kick motion upon activation. 
This coordinated deformation converts the local contraction of LCEs into global architectural expansion, yielding a macroscopic response opposite to the material's intrinsic actuation.}

To experimentally demonstrate the programmable area expansion functionality of LCE structures, we fabricate LCE structures with optimized continuous director orientations (Figure~2c).
Each LCE structure is printed in two layers, with a unit cell dimension of 45~mm $\times$ 45~mm and an overall thickness of 0.6~mm.
These LCE structures are heated on a hotplate above the nematic--isotropic transition temperature to achieve the desired actuation.
To prevent out-of-plane warping of optimized designs, we placed a glass slide of size 100~mm $\times$ 75~mm on top of the sample, with 1~mm spacers to avoid compression.
To minimize the effects of Van der Waals attraction forces~\cite{wang20203DLCEmanu} and the intrinsic stickiness of LCE in the nematic state~\cite{silva2023active}, we sprinkle magnesium powder on the sample and apply oil lubricant to the hotplate.
To validate the manufacturing accessibility and robust performance of the two optimized LCE designs, we fabricate single-unit cells as well as 2 $\times$ 2 and 3 $\times$ 3 assemblies and thermally actuate them to evaluate their expansion behavior.
\textcolor{black}{Additionally, Figure~2e summarizes the expansion ratios measured across all fabricated samples. The consistent ratios obtained for assemblies of different sizes show that the target response is retained upon periodic assembly, demonstrating the ability of the TO-based design-to-fabrication workflow to program and realize periodic LCE metamaterials (see Movie~S1 and SI for the optimization histories, DIW fabrication, and thermal expansion of both designs). 
To further distinguish the potential applications of Dsg.~1 and Dsg.~2, Figure~2f presents the simulated mean absolute inter-cell traction along the left and upper periodic boundaries (see SI for details). Dsg.~2 achieves a larger expansion ratio and higher traction intensity, indicating stronger load transfer between adjacent cells, whereas Dsg.~1 exhibits more moderate expansion and lower traction intensity. These contrasting responses highlight the distinct functional regimes targeted by the two designs.}
The first design features interconnected, braided-like limbs oriented along the diagonals, faithfully preserving the optimized topology, which is presented in the first row of Figure~2g together with its original numerical design and different assemblies, with their deformed shapes at {$110\,^\circ\mathrm{C}$} presented alongside.
Upon heating, the local director contracts perpendicular to each limb, which consequently extends the diagonally aligned limbs and drives a global area increase.
This deformation is accompanied by a reduction in thickness, consistent with the incompressibility assumption, which is evidenced by the structure gradually becoming more transparent.
We quantify the expansion ratios of the single unit cell and the meta-assemblies by 
tracking the deformed positions of the perimeter nodes.
\textcolor{black}{All cases yield similar values of approximately $A/A_0=1.40$ (Figure~2e), where $A_0$ and $A$ denote the undeformed and deformed areas. These experimental measurements are in close agreement with the numerically predicted expansion ratio of 1.401, confirming that the optimized designs are readily realizable with robust performance.}

\begin{figure}
    \centering
    \includegraphics[width=0.75\textwidth]{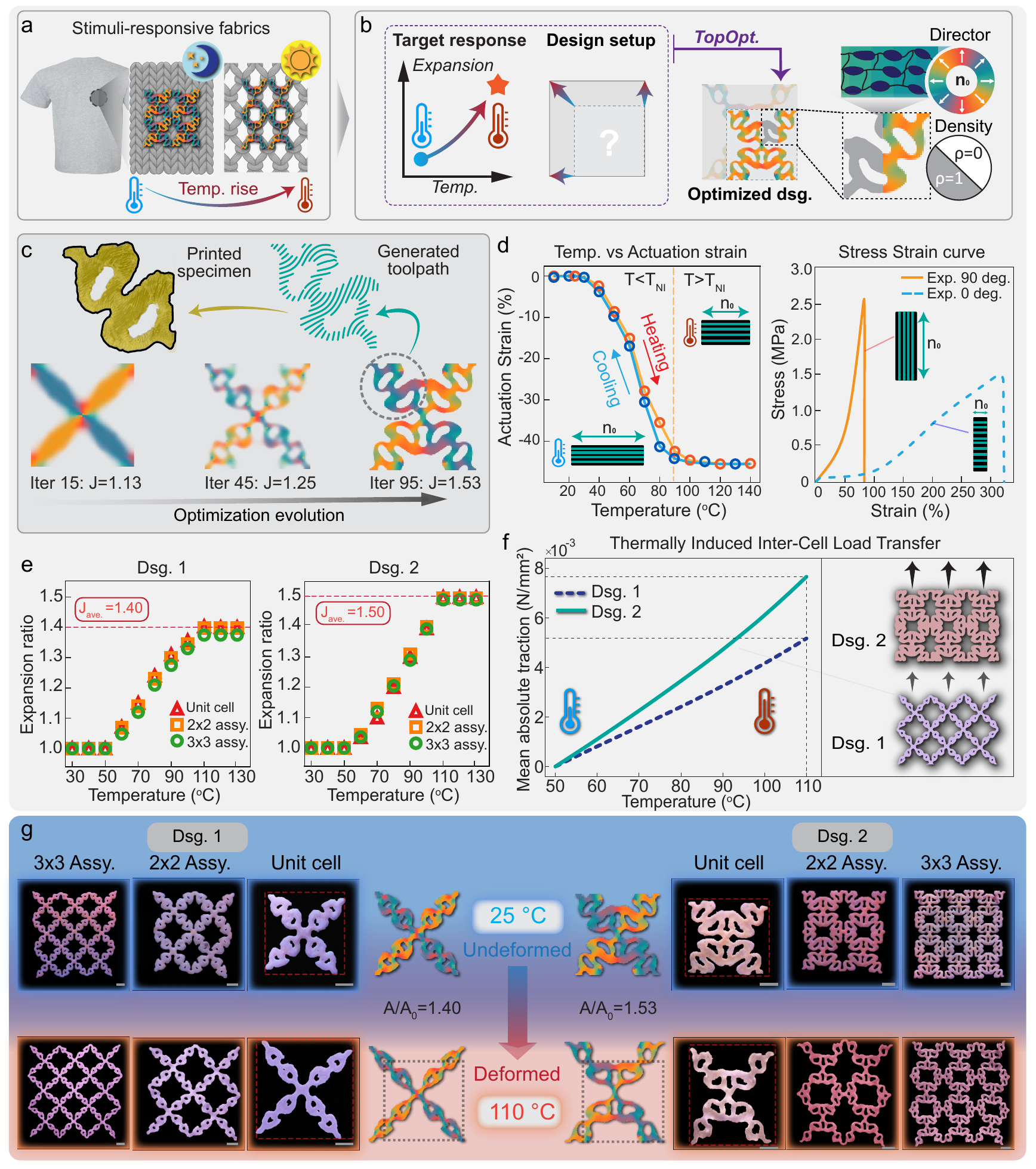}
    \caption*{%
    \textcolor{black}{\textbf{Figure~2. LCE metastructures with programmable thermally induced area expansion.}}
    \textcolor{black}{The abbreviations used in the figure are: Temp., temperature; Dsg., design; and Exp., experiment; Assy., assembly. }
    (a)~LCE structure with programmable area expansion, enabling stimuli‐responsive intelligent fabric applications.
(b)~Topology optimization framework tailored for LCEs to achieve programmable planar area expansion; each design unit is optimized for both geometric features and director orientation patterns.
(c)~Design evolution of Dsg.~2, showing representative configurations across iterations. DIW toolpaths are generated from the optimized configuration at iteration~95, and the sample is subsequently fabricated using DIW.
\textcolor{black}{(d)~Thermomechanical characterization of uniformly aligned LCE strip specimens. Left: temperature-dependent spontaneous actuation strain of a specimen with horizontally aligned directors ($0^\circ$), showing recovery of its undeformed configuration upon cooling. Right: tensile stress--strain responses of LCE strips with director orientations of $0^\circ$ and $90^\circ$ under vertical uniaxial tension.}
(e)~Area expansion ratio versus temperature for Dsg.~1 and Dsg.~2, along with the expansion ratios of their corresponding $2\times2$ and $3\times3$ assemblies.
\textcolor{black}{(f)~Temperature–traction relationship for Dsg.~1 and Dsg.~2 during thermal actuation, obtained from numerical simulations, indicating potential suitability for different application scenarios.}
(g)~Undeformed and actuated configurations of Dsg.~1 and Dsg.~2. Experimental and simulated single‐unit deformations show close agreement, and the experimental deformations of their $2\times2$ and $3\times3$ assemblies are shown alongside, the scale bar represents 10 mm for all samples.
    }
    \label{fig:2D_Expansion}
\end{figure}

Unlike the first design, which primarily relies on local director-induced contraction to generate expansion, the second design incorporates more sophisticated, muscle-like interlinked structures that unfold and deploy under thermal actuation, thereby transforming localized contraction into global expansion through conspicuous shape morphing.
To physically realize the desired performance, we fabricate this design via DIW at the unit-cell, $2 \times 2$, and $3 \times 3$ meta-assembly scales (Figure~2g),
faithfully preserving the optimized features. 
Across all architectures, the characteristic "muscle" segments unfold during actuation which resembles a frog kicking its leg to leap, highlighting a striking convergence between biological motion and optimally engineered mechanical architectures. 
\textcolor{black}{The expansion ratio of Dsg.~2 is measured to be approximately 1.5 across the fabricated assemblies, in close agreement with the numerically predicted value of 1.534. The remaining relative error of approximately $6 \%$ may arise from friction with the heating platform and mechanical constraint imposed by the overlying glass slide. The greater expansion achieved by Dsg.~2 relative to Dsg.~1 broadens its potential for applications such as deployable heat-shielding architectures.}
Owing to the automatically engineered delicate geometry and director patterns, our framework produces structures that realize counter-intuitive global expansion in opposition to the intrinsic thermal contraction of LCEs—behavior unattainable through empirical intuition or director programming alone.

\subsection*{On-demand shape morphing of LCEs into arbitrary-form complex surface}
\label{subsec:surfacefitting}

Active shape morphing from planar sheets into specified surfaces is desired for applications such as flexible electronics and in-vivo medical devices~\cite{zhao2025spatiotemporally}.
Accurate programming of LCE shape morphing into complex target surfaces with many details (such as a human face) is highly challenging, which requires delicate manipulation of LCE director distributions while retaining fabricability.
Such a challenge can be readily addressed by our proposed inverse design framework.
\textcolor{black}{Although prior studies have used LCEs to achieve high-fidelity surface morphing~\cite{zhaiDigitalProgrammingLiquid2022,aharoniUniversalInverseDesign2018}, this work explores the integration of element-level inverse design with DIW-based simultaneous LCE deposition and alignment. This approach can be adapted to different target geometries within the same computational framework while retaining the practical advantages of DIW fabrication.}

To demonstrate, we aim to achieve LCE shape morphing into the shape of a human face by optimizing the director distribution. We use two different human faces as target surfaces, as shown in Figure ~3a and b, respectively. The first one (Dsg. 1) represents a general and relatively smooth face, while the second one (Dsg. 2) features more details such as a mustache and a rounder nose tip, making it more challenging for accurate shape programming. The surfaces are defined through the level-set function $g(\mathbf{x}) + H = 0$, where $g: \mathbb{R}^3 \rightarrow \mathbb{R}$ describes the surface and is obtained by sampling points on it, $\mathbf{x}$ denotes the spatial coordinates, and H is a scalar dummy design variable used to control the vertical position of the surface (see Subsection 3.2 of SI for more details). 
\textcolor{black}{
Starting from an initially uniform director field embedded within a solid rectangular domain, intricate director patterns progressively emerge around prominent facial features, particularly the nose tip and eye sockets. The optimized bilayer director fields of the two inverse designs are shown in Figure~3c and Figure~3d, yielding fitting errors of $3.22 \%$ and $6.07 \%$ between the simulated deformed configurations and their respective target surfaces. As a representative example, the design evolution and fitting-error history of Dsg.~2 are presented in Figure~3e. Large fluctuations in the objective are observed during the initial iterations, accompanied by substantial variations in the evolving director fields and deformed configurations. During these early optimization iterations, the moving asymptotes and effective move limits have not yet stabilized, permitting relatively large simultaneous updates of the design variables that, through the coupled nonlinear thermomechanical response, produce a temporarily nonmonotonic objective history. Beyond iteration 150, the objective exhibits minor oscillations and increases slightly to approximately $6 \%$. This behavior arises from the adaptive increase in the norm parameter within the fitting-error formulation, which progressively places greater emphasis on localized regions with larger fitting errors. The slight increase in the objective therefore reflects this stronger weighting rather than a deterioration in the overall shape-matching performance ({for details, see SI, Section~3.2}). The optimization of Dsg.~2 converges after 190 iterations with a fitting error of $6.07 \%$, and both optimized designs accurately reproduce their respective target surfaces.
}{For both designs, the directors in the two layers exhibit highly complex and distinct patterns required to accurately morph into the target face profiles, with rapidly varying and non-intuitive director orientations in regions involving larger out-of-plane bulging or more complex deformation modes, such as the nose tip and eye sockets.}
Despite their complexity, the optimized director patterns are smoothly converted into DIW-printable toolpaths via the proposed path-generation algorithm. The DIW-printed specimens, with a size of 60 mm × 40 mm × 1 mm for Dsg. 1 and 50 mm × 50 mm × 1 mm for Dsg. 2, are shown at the bottom of Figures 3c and 3d (see Movie S2 and SI for the FEA setup, iterative design evolution during optimization, toolpaths generation, DIW manufacturing, and simulated actuated deformation).

\textcolor{black}{
For experimental validation, wooden clips were applied at the midpoints of the top and bottom edges of each DIW-printed LCE sheet, clamping small regions symmetrically about the central axis on both the front and back surfaces. This clamping configuration reproduces the numerical boundary conditions illustrated in Figure~3a ({for details, see SI, Section~3.2}). The clamped samples were subsequently actuated by heating in an oven.}
The deformed shapes of the two designs are shown in Figures~3f and g, respectively, along with their numerical counterparts.
Upon heating, the central region of Dsg. 1 warps up to form the nose while the regions above the nose bend downward to create the sunken eye sockets, which collectively morph into the target human face, as shown in Figure 3f. The specimen's deformed shape matches well with the numerical simulation. 
For Dsg.~2, the more complex LCE distribution produces richer facial details, including the eyes, nose, and mustache, as shown in Figure 3g, and the experimentally actuated shape agrees well with the simulation. The newly generated mustache features are highlighted in both the experimental and numerical deformed surfaces, corresponding to the more complex director variations that emerge in the first layer compared with Dsg.~1, as shown in Figure 3c,d. These sharp variations in spatial director orientations lead to neighboring directors contracting along different directions, collectively driving out-of-plane protrusions that generate prominent facial features.
These results further demonstrate a level of compatibility with diverse target geometries that is difficult to achieve through purely intuitive trial-and-error design, allowing the framework to generalize to sophisticated morphing surfaces by simply changing the target surfaces (see Movie~S2 for the experimental actuation of Dsgs.~1 and 2, \textcolor{black}{See Figure~S2c for a quantitative comparison between the numerical and experimental deformation}). 
\textcolor{black}{Additionally, both samples fully recover their initial flat configurations upon cooling and maintain stable actuation over 40 heating--cooling cycles, consistently reproducing their respective target features and thereby demonstrating the functional repeatability of the fabricated LCE structures.}

\begin{figure}
    \centering
    \includegraphics[width=0.75\textwidth]{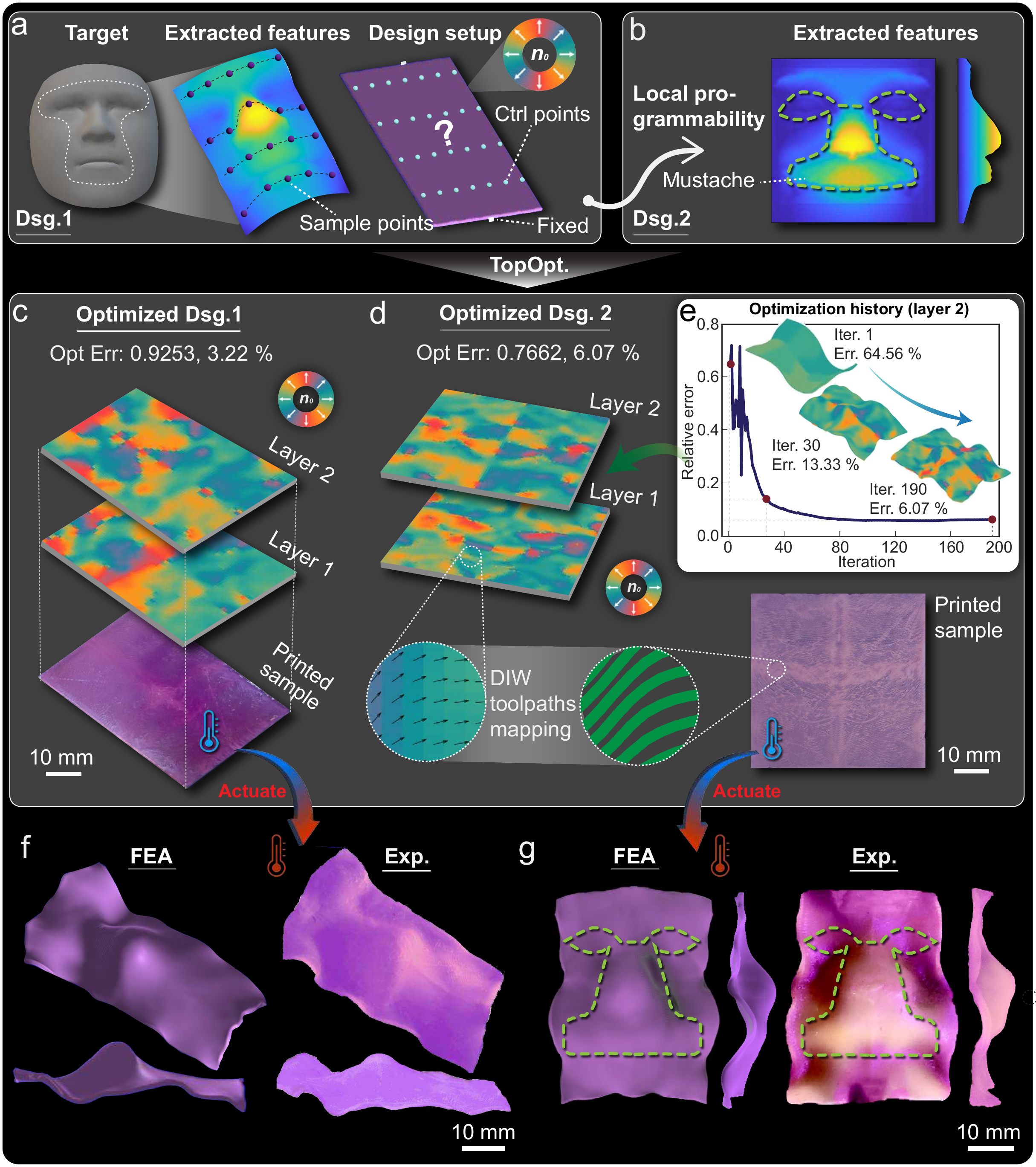}
    \caption*{%
    \textbf{Figure~3. On demand shape morphing of optimized LCE sheets into multiple target human faces.}
    \textcolor{black}{The abbreviations used in the figure are: Ctrl., control; Dsg., design; Opt. Err., optimization error; and Exp., experiment.}
    (a)~Extraction and sampling of critical facial features from the target human face to define the objective surface for the LCEs optimization. Control points are distributed on the undeformed LCE sheet to minimize fitting errors after deformation.
(b)~For localized feature programming, more distinctive attributes—such as a mustache and a rounder nose tip—are extracted from a second, more characteristic target face to form a second objective surface.
(c)~Optimized bi-layer LCE design that deforms to reproduce the first target facial surface. Both layers contain spatially varying director patterns, and the physical LCE sample is fabricated using DIW based on the generated printing toolpaths.
(d)~Second optimized LCE design that reproduces the more distinctive facial features of the second target. DIW toolpaths are generated directly from the numerical design, and the insets confirm agreement between the optimized director orientations and the DIW toolpaths alignment directions.
(e)~Optimization history of the second design, showing the decreasing normalized fitting error, with three inset configurations illustrating representative deformed designs from different iterations.
(f)~Deformation of optimized Dsg.~1 from both experiments and simulations, shown from a 3D top view (top row) and a side view (bottom row), demonstrating close agreement between the two.
(g)~Deformation of optimized Dsg.~2 from FEA simulations and experiments, shown from top and side views. The green dotted line highlights characteristic features—such as the rounder nose tip and mustache—unique to the second target, further confirming consistency between numerical predictions and experimental results.}
    \label{fig:humanface}
\end{figure}

\subsection*{Autonomous thermo-rotational LCE actuator realizing accurate trajectory}
\label{subsec:rotationcontrol}

The inverse design also allows us to precisely manipulate LCE structure's locomotion, which could benefit applications such as active wound closure and circuit switch (Figure~4a). We demonstrate this by programming a thermally actuated self-rotation of an LCE structure as illustrated in Figure~4b, where both the geometry and director orientation of the LCE structure are optimized simultaneously.
\textcolor{black}{By defining the objective function as the maximum normalized error between the actual and the target locations of control points at the four corners of the LCE sheet over the process of continuous temperature rise. Mathematically, the objective function is 
$ 
\min_{\rho,\xi} \max_{s=1}^{n_{\mathrm{ctrl}}} 
\sum_{i=1}^{n_{\mathrm{DOF}}} \frac{(u_i^{(s)} - u_i^{*(s)})^2}{(u_i^{*(s)})^2},
$ 
where $\rho,\xi$ are designable pseudo density and LCE directors orientations respectively, additionally $u_i^{(s)}$ and $u_i^{*(s)}$ are actual and target displacement for the $i-$th control DOF at the $(s)$ heating stage (for more details, see Subsection 3.3 of SI).}

Regarding the target trajectory, circular-connected rotational traces were stipulated to achieve the target trajectory with a maximum rotation angle of $63^\circ$ (Figure 4b). 
The central $2\,\mathrm{mm} \times 2\,\mathrm{mm}$ region of the design domain was fixed to provide an anchoring axis, while four-fold rotational symmetry with respect to the quadrants was imposed to guide the optimizer toward an efficient solution.
Using this configuration, we generated a family of four bi-layer optimal LCE structures (denoted as Dsg. 1 to Dsg. 4, respectively in Figure 4c), 
all capable of achieving the target rotational deformation while exhibiting distinct design and mechanical features such as different overall stiffness. 
As the most compliant design, Dsg. 1 generates a central windmill-like anchoring region with four slender extruded bars connecting to primary rotating blades.
The overall design also incorporates weight-reducing hollowed regions to facilitate rotation and sharp ridges that transmit motion to the corner output DOFs. 
Moreover, the bi-layer director parameterization assigns distinct director distributions to each layer within the optimized geometry, enabling precise rotation control via a continuous director pattern compatible with DIW fabrication.
As the stiffness increases, the diagonally aligned cross structure evolves to generate larger reaction forces during rotation, and the hollowed regions shrink to further stiffen the design.
These trends become prominent from Dsg. 1 to 4 (Figure 4c), 
and their reaction force–displacement responses under mechanical loading confirm the increased stiffness (Figure 4h). 
The numerical rotational deformation of Dsg. 1 is shown in Figure 4d, 
with the control-point trajectory overlaid on the target path.
The optimized design maintains consistently small fitting errors across four control stages, accurately achieving the prescribed rotational performance and demonstrating the capability for multi-stage, complex kinematic control of LCEs that is essential for advanced dynamic behaviors  (see Figure S1 and Movie S3 for design evolution, the simulated deformations of Dsgs. 1-4).

A quadrant of the DIW toolpaths generated for the optimized LCE topologies is schematically illustrated in Figure 4f, 
indicating continuous filament deposition.
Using these high-quality toolpaths, a rotator sample with the same dimensions as the numerical setup was fabricated employing four printed layers of 0.25~mm thickness, with every two printed layers constituting one design layer, yielding an overall size of 60~mm (height) $\times$ 60~mm (width) $\times$ 1~mm (thickness) as shown in Figure 4e.

To evaluate the experimental performance of the fabricated sample, we fixed the central anchoring region and actuated the sample with a hotplate. 
Upon heating, the sophisticated director patterns within the four-fold "windmill" blades are actuated, transmitting forces through the four slender bars to counteract the reaction forces from the central anchoring region, collectively producing a moment that drives the entire structure to rotate along the prescribed trajectory (Figure 4d; 
see Movie S3 and SI for the experimental deformation of Dsg. 1).
We quantified the rotation by tracking the deformed locations at the control DOFs and comparing them with numerical results (Figure 4g). 
\textcolor{black}{The sample achieves a rotation angle of approximately $59.7^\circ$, closely matching the numerically predicted value of $62.7^\circ$ and corresponding to a relative error of $4.7\%$. This discrepancy could be further reduced by improving fabrication precision and minimizing friction between the sample and the heating platform.}
Along with consistently negligible fitting errors across the four control stages, the physical samples reproduced the overall numerical deformation trajectories, demonstrating the effectiveness of our methodology in achieving precise kinematic control by controlling the motion at key stages.
The fabricated sample also maintained similar intact rotational deformation over more than 30 actuation cycles, showing the 
cycling robustness of the printed structure.

\begin{figure}
    \centering
    \includegraphics[width=0.75\textwidth]{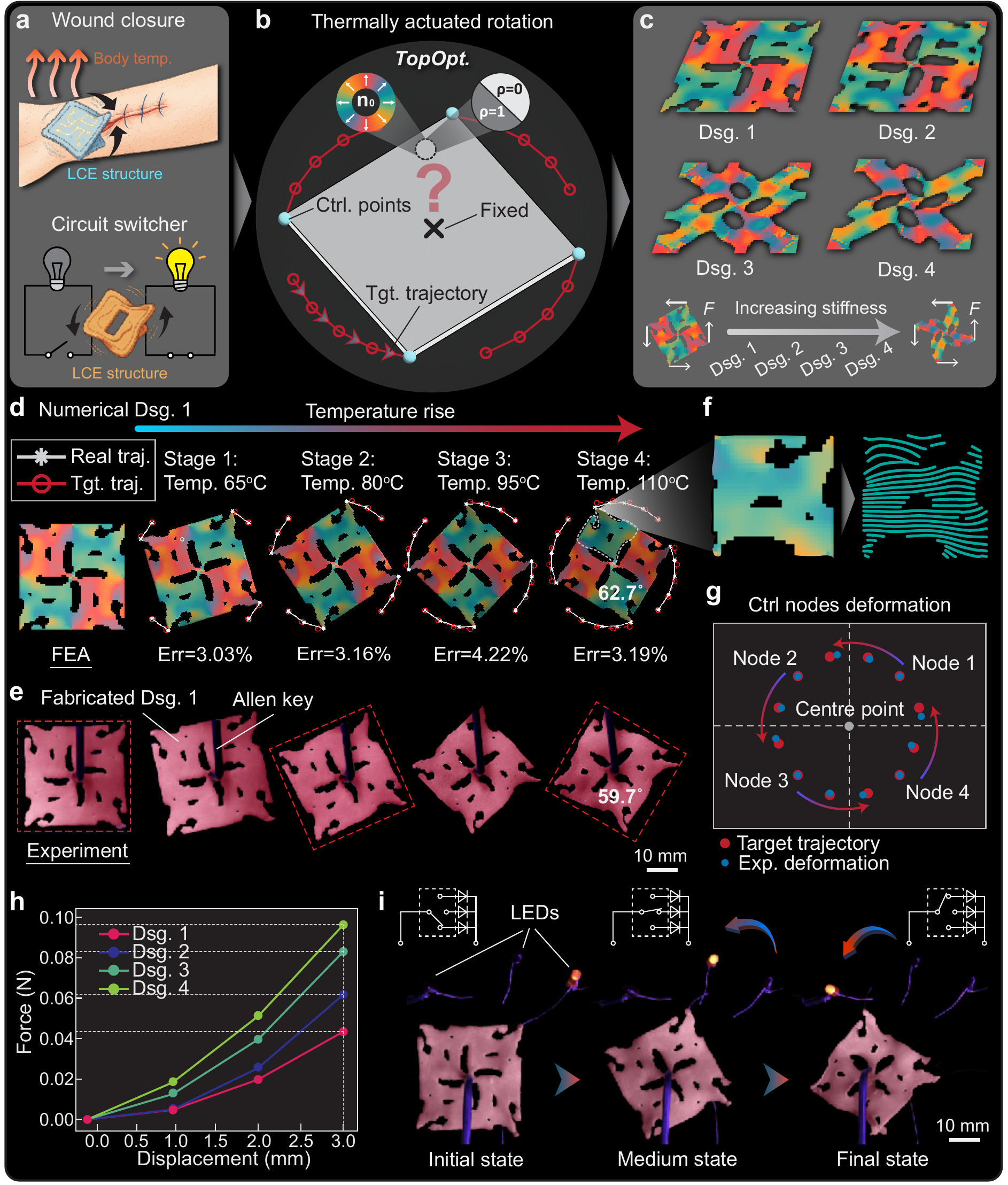}
   \caption*{%
\textbf{Figure~4. Topology-optimized LCE actuator with regulated and autonomous thermo-rotational trajectory control.}
\textcolor{black}{The abbreviations used in the figure are: Ctrl., control; Dsg., design; Tgt. traj., target trajectory; and Exp., experiment.}
(a) Representative application scenarios, including an LCE-based rotational wound-closure stitcher and a rotational circuit switcher, motivating the development of controlled LCE rotators.  
(b) Design domain for topology optimization targeting prescribed rotational trajectories, showing the target curve, control points, and the central fixed region. 
(c) Increasing the stiffness requirements yields four distinct optimized designs, generating stronger forces under displacement loading; for each, two continuous director-orientation layers are embedded within the optimized geometry.    
(d) Simulated thermally induced deformation of optimized rotator Dsg.~1, showing deformed configurations at multiple loading states. Achieved trajectories closely match the targets, and the final rotation angle is reported.  
(e) Experimental deformation of the fabricated Dsg.~1, with an Allen key constraining the center to replicate numerical boundary conditions. Top-view images at key heating states agree closely with simulations, validating the design framework; the final rotation angle is also measured.  
(f) Printing toolpaths for one quadrant of Dsg.~1, illustrating path alignment with optimized director orientations and geometry.  
(g) Projected control-node positions at various actuation stages, demonstrating close agreement between achieved and prescribed trajectories.  
(h) Force–displacement curves from independent mechanical loading tests (displacement applied at four corners) for Dsg.~1–4, confirming the monotonic stiffness increase across designs.  
(i) Demonstration of Dsg.~1 as a single-pole triple-throw (SPTT) switch: LEDs placed at different locations along the deformation trajectory make sequential contact with the rotator's attached wires during actuation, illuminating in a timed sequence as the structure rotates.
}
    \label{fig:rotator}
\end{figure}

We further validated the potential of the designed LCE rotator for practical soft-electronics applications.
As shown in Figure~4i, 
we constructed a DC parallel circuit in which the LCE rotator serves as a single-pole triple-throw (SPTT) switch.
Conductive thread was attached to the back of the LCE sheet to minimize mechanical interference during rotation~\cite{chen2024knotted}.
Three light-emitting diodes (LEDs) were positioned along the rotational path and were sequentially illuminated as the rotator deformed following the prescribed trajectory, thereby connecting different circuit branches during heating.
The three LEDs lit in chronological order from left to right, corresponding to the undeformed, intermediate, and fully actuated angular positions (Figure~4i; see Movie S3 and SI for the experimental deformation of LCE rotator as a circuit switch to illuminate LEDs). 

\subsection*{Bio-inspired LCE soft robots with multifunctional actuated behaviors}
\label{subsec:Octopus inspired multi functional robotic arm}

Nature offers a rich source of biological inspiration for engineering LCE, and prior studies have demonstrated bio-inspired LCE designs, including a cephalopod-inspired jet swimmer~\cite{chenScalableJetSwimmer2025} and a compliant gripper actuated by vascular LCE muscles~\cite{heBioinspiredDesignVascular2019}.
Inspired by the dexterous arm movements of cephalopods (e.g., octopuses and cuttlefish) during object grasping and by the "front crawl" motion of human freestyle swimming, we develop an octopus-arm-shaped LCE soft actuator with dual functional objectives.
\textcolor{black}{Through tailored topology optimization, the same LCE-arm design supports both object gripping and non-monotonic propulsion-inspired motion, implemented using distinct actuation setups.
The proposed bi-functional compliant LCE actuator provides a platform for future electrically controlled soft robotic systems operating in confined amphibious environments.}

For the octopus-arm LCE design aimed at achieving dual functions of grasping and flapping, we introduce a two-stage trajectory to regulate the motion of the compliant LCE arm (Figure 5a) 
and applied the same optimization objective function used for the rotator design to achieve the desired LCE motion.
To emulate the hydrodynamically efficient contour of an octopus arm, we fix the geometry of a single arm and optimize only the incorporated director pattern within this predefined shape.
In the first stage of the predefined trajectory, target locations are designed to induce contraction of the LCE structure, which enables the arm to clench around an object once multiple symmetric arms are positioned in a circular configuration. 
When switched for flapping motion, the same contraction serves as the initial movement for generating a crawling or hugging motion against the surrounding fluid, effectively preparing the arm for the second-stage propulsion. 
In the second stage of motion, the arm tips slide downward, stretching the previously bent structure.
\textcolor{black}{This motion is designed to emulate a propulsion-inspired stroke that would displace the surrounding fluid downward, although actual fluid displacement and thrust generation are not evaluated here.
}
In addition, in gripper applications, this stage reinforces the grasp by further tightening around the target object while maintaining positional stability.
By composing this two-stage, non-monotonic deformation trajectory (Figure 5a), 
we guide the optimizer to iteratively obtain the desired bi-functional grasping–flapping LCE design.
\textcolor{black}{Starting from an initially uniform director field oriented at $90^\circ$, the optimized bifunctional LCE arm shown in Figure~5b is obtained after 105 optimization iterations (Movie~S4).
Inside the prescribed hydrodynamically efficient exterior inspired by the octopus arm, two layers of sophisticated director patterns are smoothly fashioned to fulfill the design objectives while ensuring compatibility with DIW. }
The optimized design is predominantly distributed with $90^{\circ}$-aligned directors in both layers, which, upon actuation, stretch the structure along the central axis while the wider end of the arm remains fixed.
Additionally, to enable the initial crawling behavior, distinct director patterns are generated in the middle region, which act in concert with the primary stretching zones to induce out-of-plane warping.
To validate the functional performance of the physical design, we generated the corresponding printing toolpaths from the optimized design, and fabricated the LCE actuator via DIW (Figure 5c). 
The sample was suspended vertically to minimize gravitational effects, and clips were used to clamp the wider top edge, replicating the numerical boundary condition.
The simulated and experimental deformations of the optimized LCE actuator are illustrated in Figure 5d, 
showing their close agreement at two critical control states, confirming that the optimized bio-inspired LCE actuator successfully achieves the target non-monotonic "crawling–propelling" motion.

Upon actuation using an IR lamp, the sample exhibited two-stage deformation as illustrated in Figure 5d. 
In the first stage, the LCE director contracts, resulting in bending of the octopus arm in regions near both the bottom tip and the top, preparing the body to generate the subsequent propelling motion. 
{During the second stage, further heating drives the arm downward from its first-stage curled configuration, thereby completing the propulsion-inspired stroke.}
The second-stage motion is driven by the progressively actuated contraction of the directors in the upper portion of the octopus arm.
\textcolor{black}{A comparison between the deformed positions of the control nodes and their numerical counterparts is presented in Figure 5e, 
demonstrating good agreement between deformed configurations (see Movie S4 and SI for the DIW fabrication and design deformation).
For the tip control node, the relative displacement errors between the FEA predictions and experimental measurements are $4.26\%$ at stage~1 and $2.89\%$ at stage~2, indicating close quantitative agreement. The corresponding errors for the mid control node are larger, reaching $31.5\%$ and $21.7\%$, respectively. These larger errors primarily reflect the smaller displacement magnitude of the mid node, which amplifies deviations when expressed in relative terms.
}
\textcolor{black}{Although the optimized LCE arm realizes the prescribed non-monotonic, propulsion-inspired trajectory upon heating, it has not yet been tested underwater; its ability to generate thrust and enable soft-robot locomotion therefore remains to be established. 
Extending this design to underwater soft-robotic systems will require validation of repeatable actuation under fluid conditions and consideration of coupled fluid--structure mechanics, which constitute important directions for future work.}

\begin{figure}
    \centering
    \includegraphics[width=0.85\textwidth]{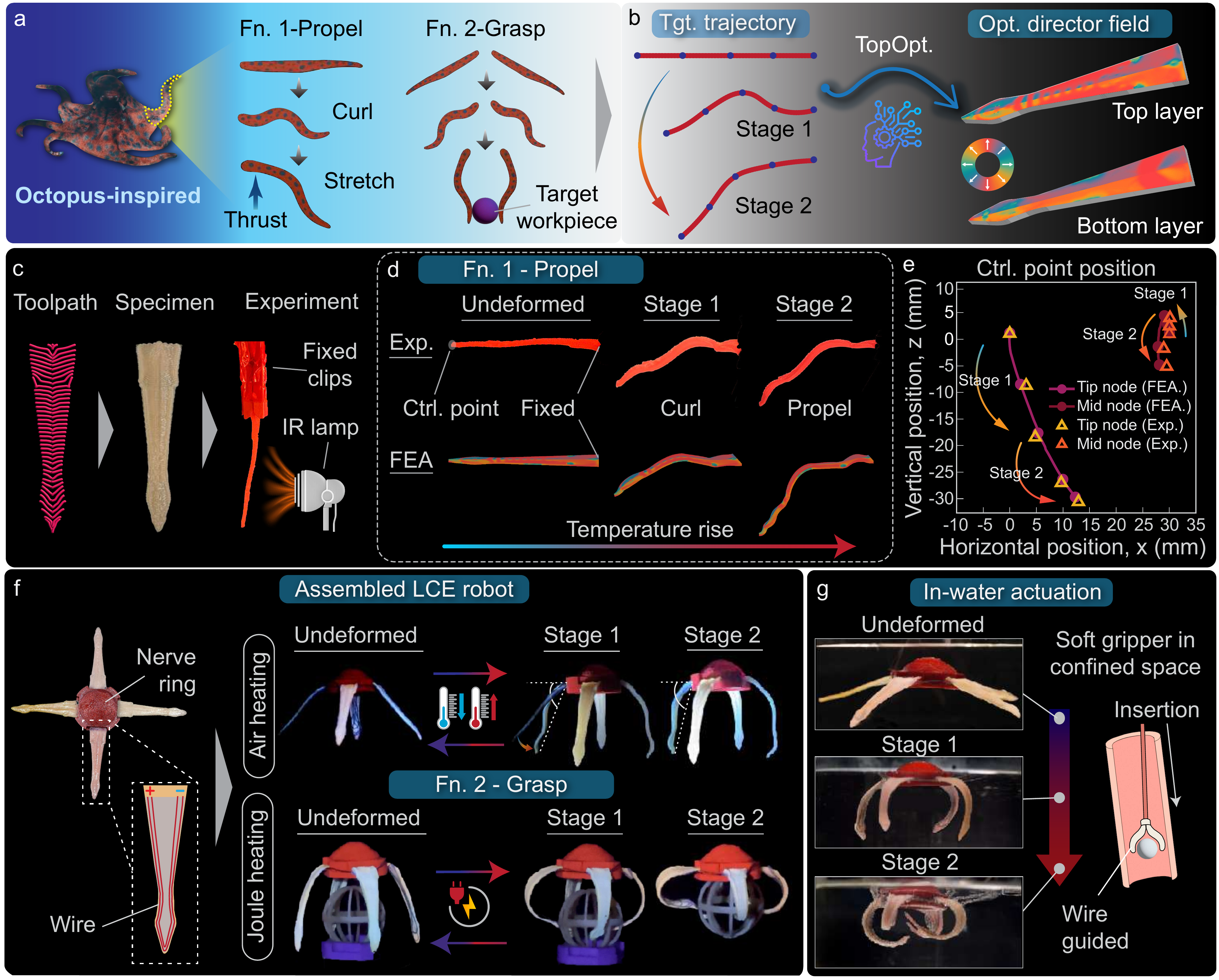}
    \caption*{%
\textbf{\textcolor{black}{Figure~5. Bio-inspired LCE soft robots with multifunctional actuated behaviors.} }
\textcolor{black}{The abbreviations used in the figure are: Fn., function; Tgt. , target; Btm., bottom; Exp., experiment; Ctrl. pt., control point; }
(a)~Conceptual illustration of the multifunctional LCE actuator design, inspired by the dexterous motion of cephalopods to enable multiple functionalities.
(b)~Target trajectories for the two-stage "crawling–propelling" motion, along with the optimized bi-layer LCE design shown on the right integrating shape of an octopus arm.
(c)~Generated DIW toolpaths based on the optimized design and the resulting DIW-fabricated LCE soft actuator. During thermal actuation tests, the actuator is suspended vertically to minimize gravity effects, with one end fixed to reproduce the boundary condition and an infrared (IR) lamp used for heating.
(d)~Actuated deformation of an individual LCE actuator (top) compared with FEA simulations (bottom), showing close agreement. Two representative control states highlight the non-monotonic deformation modes realized during heating, with bending for crawling followed by stretching for propulsion.
(e)~Comparison between simulated and experimental deformation trajectories of the tip and mid control nodes, projected onto the x–z plane. The close agreement validates faithful reproduction of the high-fidelity simulation, and the non-monotonic upward–downward motion of the mid node confirms successful realization of the bending–stretching transition.
(f)~Four-arm assembly connected to a central dome, with heat-resistant embedded wires enabling Joule-heating actuation. The first row shows air-heated deformation states, and the second row shows electrically actuated deformation during a gripping demonstration. Upon powering the embedded wires, each arm contracts toward the central connector, allowing the assembly to successfully grasp a lattice-ball object.
(g)~Experimental demonstration of the second functionality-underwater curling. Joule-heating actuation induces the desired curling motion when placed underwater, indicating potential for soft-gripper operation in confined amphibious environments.}
    \label{fig:octopus}
\end{figure}

Next, we designed an octopus-inspired assembly featuring multiple arms to achieve a simultaneous two-stage flapping motion.
Four arms were fabricated and glued onto a FDM printed central nerve ring (Figure 5f). 
\textcolor{black}{The octopus assembly was suspended in a vacuum oven to ensure uniform heating of all four arms; consequently, the arms were initially not perfectly horizontal because of gravitational sagging.}
During the experiment, the oven was gradually heated to 110 $^\circ\mathrm{C}$ allowing simultaneous crawling/propelling of the four octopus arms. 
In the first stage, the gradual bending of sample around tip and the upper regions is observed, closely resembling the deformation of the aforementioned individual arm.
In the second stage, all arms perform a propulsive motion inwards towards the center of nerve ring, mimicking the swimming stroke akin to single arm actuation. 

\textcolor{black}{The arm tips curl inward, undergoing displacements of approximately $26.8\,\mathrm{mm}$ from their initial positions. This two-stage response produces a propulsion-inspired flapping stroke reminiscent of swimming. However, the non-monotonic trajectory is less pronounced in the assembled device than in the single-arm demonstration because gravitational sagging alters the intended horizontal initial configuration. Developing a stiffer architecture that resists gravitational sagging while faithfully preserving the programmed actuation represents an important direction for future application-oriented designs.}

Furthermore, we effectively demonstrate the grasping functionality of the generated LCE soft robots, which requires more pronounced curling deformation of each arm.
Specifically, stainless-steel conductive threads were bonded to the back of the LCE arms and connected to a central nerve ring, and these threads were wired to a power source to generate Joule heating upon activation (Figure~5f; 
Movie~S4), yielding localized, higher-intensity Joule heating of individual arms.

\textcolor{black}{Additionally, the glued conductive wires stiffen the LCE arms and suppress the downwards propulsion motion, which alter the prescribed two-stage non-monotonic trajectory. However, the added stiffness provides the mechanical support required for grasping, while localized Joule heating induces sufficient curling for the arms to contact, close around, and securely hold the target object. The gripping function is therefore demonstrated using a distinct experimental setup that leverages the multistage actuation mechanism, rather than reproducing the prescribed trajectory with the accuracy achieved in the isolated-arm validation.}
As illustrated in Figure 5f, 
the actuated four arms simultaneously curve inwards to grasp the ball on the supporting platform. 
During the curling motion, the arms remain in contact with the ball, clamping and lifting the object and maintaining the grasp throughout actuation.
To further showcase the underwater grasping potential of this gripper, we submerge the octopus assembly inside the water (Figure 5g). 
Upon actuation, the arms curl inward synchronously to achieve pronounced grasping, demonstrating effective amphibious gripper functionality without performance degradation. 
Despite faster thermal dissipation under water, the gripper maintains performance comparable to the in-air actuation, attributable to the improved electrically induced Joule-heating method.
This robustness across different environments highlights the potential of the optimized bi-functional LCE designs as remotely controllable soft robots for operation in confined amphibious environments.
Moreover, the maintained in-water actuation reveals a promising avenue for novel underwater locomotion strategies to be explored in future studies.

\section*{DISCUSSION}
\textcolor{black}{
This work presents a multilayer co-optimization framework and monolithically integrates it with an end-to-end fabrication pipeline, enabling the precise programming and physical realization of diverse thermally activated 3D responses in freeform LCE architectures. Built upon a rigorous non-ideal LCE model, the multiphysics framework autonomously co-optimizes structural geometry and layer-specific director distributions while promoting compatibility with DIW-based 4D printing. Compared with inverse-design approaches that employ reduced-order representations~\cite{wangAutonomousCodesignFabrication2025,fuchiTopologyOptimizationDesign2015} or focus on two-dimensional planar structures~\cite{liComputationalMorphogenesisLiquid2024}, the full-scale nonlinear hyperelastic FEA and multilayer co-optimization framework here incurs a higher computational cost that increases with mesh resolution. This additional cost, however, provides greater physical fidelity and expanded design freedom, which enable the simulation and inverse design of more complex three-dimensional deformations. A practical balance between computational scalability and design resolution can be achieved by adjusting the mesh resolution while accounting for the printing resolution and prescribed minimum length scale. Several modeling limitations nevertheless remain. The present formulation assumes frozen directors and does not account for viscoelasticity~\cite{jiang2025viscoelastic,chehadeFiniteElementModeling2024}. Extending the constitutive model to incorporate viscoelastic effects and independent director rotation would enable the explicit treatment of time-dependent network deformation and director reorientation. 
The quasi-static treatment of programmed locomotion may limit the framework's applicability to rapidly actuated soft-robotic systems, motivating the incorporation of dynamic simulations in future work.
Gravity and self-contact are neglected when programming deformation trajectories, potentially contributing to discrepancies between simulations and experiments for structures undergoing extreme deformation. Numerical regularization strategies that penalize higher-order deformation modes~\cite{frederiksen2025improved} could also improve stability in the presence of severe mesh distortion. Incorporating these physical effects and stabilization strategies, together with more rigorous material-parameter calibration~\cite{wei2025synthesis}, would further enhance the predictive fidelity and numerical robustness of the framework.
}

\textcolor{black}{
At the formulation level, distinct functional objectives---including surface fitting, programmable expansion of assemblies from local contraction, and multistage trajectory control during thermal activation---are incorporated into a unified topology-optimization framework. 
Although surface fitting has previously been achieved successfully through curvature mapping~\cite{aharoniUniversalInverseDesign2018,zhaiDigitalProgrammingLiquid2022}, here it is realized through a control-point-based objective formulation and directly translated into DIW-fabricated LCE architectures for simultaneous LCE deposition and alignment. The multilayer optimization formulation is directly integrated with a continuous director-filtering scheme to regularize the layer-specific orientations and generate smooth director fields compatible with layer-wise DIW, while subsequent post-processing strategies accommodate more rapidly varying director patterns. 
Manufacturability of the designs could be further strengthened by directly incorporating constraints on filament overlap, interlayer sagging, and filament spacing into the optimization so as to mitigate fabrication-induced imperfections.
Exploring level-set-based parameterizations~\cite{fernandezOptimalDesignFiber2019} may additionally provide a route toward smoother director fields and curvature-controlled toolpaths.
}

\textcolor{black}{
The proposed topology-aware toolpath-generation algorithm with an advanced nonlinear formulation seamlessly converts optimized LCE geometries and complex director distributions into smooth, well-planned, executable DIW toolpaths. The resulting autonomous design-to-fabrication strategy enables the realization of diverse user-prescribed LCE responses. Its manufacturing robustness is supported by the successful fabrication and experimental validations of LCE metastructures exhibiting programmable expansion despite local director contraction, morphing sheets that capture prescribed features of complex three-dimensional target surfaces, self-rotating compliant actuators governed by prescribed trajectories, and multifunctional octopus-arm actuators exhibiting propulsion-inspired trajectories and electrically driven gripping. The printability of the optimized director fields is physically demonstrated by extrusion in the nematic state, which induces mesogen alignment along the printing direction through shear. Representative polarized optical microscopy (POM) measurements confirm local director alignment (Fig. S5), whereas full-field characterization of the complete spatial director distribution remains an important future direction. 
The toolpath-generation procedure could also be augmented with global path-sequencing methods, such as traveling-salesman-problem strategies~\cite{applegate2003implementing}, to limit unintended material deposition during internal traversals, and improve DIW efficiency. 
Extending planar DIW to embedded 4D printing~\cite{mcdougall2023free} could enable spatial programming of director orientations in self-supporting three-dimensional architectures, expanding the accessible design space to functional $3\mathrm{D}$-to-$3\mathrm{D}$ transformations.}
Collectively, these findings unveil a promising avenue for next-generation LCE engineering that is function-generalizable, DIW-accessible, and practically reproducible.
By enabling programmable, freeform, and manufacturable LCE architectures with diverse, application-oriented thermomechanical responses, the proposed framework provides a versatile foundation for future developments in applications, such as soft robotics, wearable devices, and adaptive electronics.

\section*{METHODS}

\paragraph{Modeling and numerical simulation of LCEs} 
To model the complex nonlinear behavior of LCEs upon thermal actuation, we adopt a continuum LCE model based on the nonideal model (see Subsection 1.1 of SI for details).
Through the initial and actuated orientation order parameter that depend directly on the initial and current temperatures, the developed model captures LCE’s thermal actuation effectively.
We then employ nonlinear finite element analysis (FEA) to solve the elastostatic problem of LCEs.
We utilize the standard displacement-based 2D\&3D FEM with a total Lagrangian formulation, employing bilinear quadrilateral elements for 2D cases and trilinear hexahedral elements for 3D cases, together with Gaussian quadrature for numerical integration.
The nonlinear problem minimizing the total potential energy with respect to displacement vector is solved using Newton–Raphson method (see Subsection 1.2 of SI for details).

\paragraph{Topology optimization design of LCEs} The established topology optimization framework adopt the density design variable $\rho$ to characterize the geometry and the angle design variable $\xi$ to characterize the directors of LCEs.
The density design variable is mapped to the physical density variable $\overline{\rho}$ through the standard density filter and smooth Heaviside projection (see Subsection 2.1 of SI for details).
The stored energy is calculated via the proposed LCE material model and interpolated based on the SIMP scheme.

For the 3D LCE applications considered in this study, we adopt a multi-layer director setup within our TO framework to expand the design space and enable sophisticated functional performance, where two independent director layers with identical geometry are defined.
Within each layer, the aforementioned continuation scheme is enforced to ensure continuous director patterns compatible with DIW printing.
The directors are constrained to the 2D in-plane form \([\xi_{x},\xi_{y},0]\), with the out-of-plane component \(\xi_{z}\) set to zero, thereby enabling layer-by-layer fabrication via DIW.
Different optimization formulations targeting versatile LCEs functionalities are proposed via distinct objectives, while the sensitivity of topology optimization is analyzed utilizing adjoint method (see Section 2 of SI for details).


\paragraph{DIW printing path generation} We employ an optimal-field-based isoline extraction method to generate the printing toolpaths of optimized LCE designs, enabling the fabrication of intricate geometries with continuously varying nematic director fields.
We construct a constrained minimization problem to yield the optimal intermediate field $\mathbf{\phi}$ whose isolines align with the orientations of the optimized LCE nematic directors.
To solve this optimization problem numerically, we reformulate it as a Lagrangian penalized function, solved using a finite element method (FEM) in MATLAB.
Once the optimal intermediate field is obtained, isolines are extracted and trimmed to conform to the optimized LCE geometry (see Section 4 of SI for details),
To convert the generated printing toolpaths into G\mbox{-}code compatible, we customize G-code files that determine (i) the appropriate extrusion ratio during nozzle motion and (ii) the spacing between adjacent filaments (see Section 5 of SI for details).

\paragraph{Synthesis of LCE ink} The LCE ink was synthesized using a solvent based approach \cite{wang2020three}.
The ink is prepared by mixing (1,4-Bis-[4-(3-acryloyloxypropyloxy)benzoyloxy]-2-methylbenzene) (RM 257) (AmBeed Inc., 99.39\%), (2,2'-(Ethylenedioxy)\allowbreak diethanethiol) (EDDET) (MilliporeSigma, 95\%), dipropylamine (DPA) (MilliporeSigma, 98\%), dichloromethane (DCM) (Millipore-Sigma, ACS reagent), (2-hydroxyethoxy)-2methylpropiophenone (Irgacure 2959, MilliporeSigma, 98\%), and Rhodamine B dye  (MilliporeSigma, 95\%).
All reagents were used as received without further purification. 
RM 257 (8.2404 g, 14 mmol) is dissolved into 50 ml of solvent DCM.
Then, chain extender EDDET (2.1876 g, 12 mmol) and catalyst DPA (0.10 g, 1 mmol) are added into the solution drop-wise.
The solution is stirred overnight at ambient temperature to facilitate Michael addition reaction.
After that, photo initiator Irgacure 2959 (0.05 g, 0.2 mmol) along with RhB dye (0.0010 g, 0.002 mmol) is added into the solution.
The solution is heated at \(85\,^{\circ}\mathrm{C}\) for 24 hours to ensure complete evaporation of solvent. 

\paragraph{4D printing (DIW) of LCEs} The LCE ink is loaded into the Hyrel KR-2 printing head syringe and heated at \(85\,^{\circ}\mathrm{C}\) for 12 hours to allow for the evaporation of bubbles.
The printing head is then loaded into the Hyrel Systems-30M printer.
The specimens were printed on a glass slide coated with liquid 3d printing adhesive, at the speed of 120 mm/min and a printing temperature of \(45\,^{\circ}\mathrm{C}\).
The non-printing (traveling) moves of nozzle are at the speed of 360-500 mm/min.
The printing path is generated using a customized g-code generator (See SI for more details).
The extruded ink aligns the director in the desired orientation due to the shear force between the nozzle and the LCE ink \cite{kotikian20183d}.
To fix the director orientation, UV light (Hyrel 365 nm) at 100\% intensity is continuously illuminated during the printing to fix director orientation.
After printing, the samples are further UV-cured for 15 minutes in a custom-built high-intensity UV lamp.
The samples are removed from the UV chamber for experimental testing.

\paragraph{Mechanical characterization and thermal actuation of LCE} Stress-strain characterization of the LCE was conducted through uniaxial tensile tests using an Instron 68SC-5 machine.
With dimensions of 30 mm $\times$ 6 mm $\times$ 0.2 mm, the printed LCEs samples are secured at their top and bottom utilizing customized 3d-printed clamps.
The attachment is done using polyimide tape \cite{wang2022anisotropic}.
The tests were performed using a 500 N load cell with a constant strain rate of $0.01 s^{-1}$. 
The stress-strain plot is shown in (Figure 2d).
To evaluate the actuation performance, bi-layer rectangular LCE samples of dimensions 15 mm $\times$ 3 mm $\times$ 0.25 mm were printed using a 0.2 mm nozzle and a printing speed of 120 mm/min \cite{kotikian20183d}.
The sample was placed on a hot plate inside a Petri dish holding silicon oil and heated up to \(120\,^{\circ}\mathrm{C}\).
Upon heating, the change of samples in length was recorded by camera, and measured using the ImageJ software.
Samples were successively actuated with an average actuation contraction of -44.5\% along the print direction.

\newpage

\section*{DATA AVAILABILITY}
\paragraph{} \textcolor{black}{
The point-cloud datasets defining the two target human-face surfaces used in the LCE shape-morphing demonstrations are available through \href{https://zenodo.org/records/22036297?preview=1\&token=eyJhbGciOiJIUzUxMiJ9.eyJpZCI6IjFiMDU1ODJhLTlhZjktNDRmYi1iYTRkLTAzZjI2MTNhMGUzOSIsImRhdGEiOnt9LCJyYW5kb20iOiI0YmJkYTE4M2E2M2NlMzY2MjNmYTgxMDc5YzU1MTU5ZiJ9.5lovA6KG9yHHaFkkK2I51GIjTn_iIXaskP3JvcIbMkAMADrIGdpg4gcZ444LNnazwVWpnMkb4I3tty0TsohT9A}{Zenodo}. 
The remaining data of this study are available from the corresponding author upon reasonable request.
}

\section*{CODE AVAILABILITY}
\paragraph{} \textcolor{black}{The computational code to generate the
LCE designs is available upon request, as it is undergoing further development for future research at the time of submission.}

\bibliographystyle{unsrt}
\bibliography{mybib.bib,ybw-lce.bib}

\section*{ACKNOWLEDGMENTS}

\paragraph{} The authors acknowledge the financial support from the U.S. National Science Foundation (NSF) CAREER Award CMMI-2047692, CMMI-2344258, and NSF Award CMMI-2505649. This work was further supported by the Air Force Office of Scientific Research under award number FA9550-23-1-0297. The information provided in this paper is the sole opinion of the authors and does not necessarily reflect the view of the sponsoring agency.

\section*{AUTHOR CONTRIBUTIONS}

\paragraph{} Y.W. and A.F.A. contributed equally to this work. Y.W., A.F.A., W.L., and X.S.Z. developed the methodology. Y.W. and W.L. developed the software. A.F.A. carried out the fabrication. A.F.A., Y.W., and W.X. performed the experiments. Y.W., A.F.A., W.L., W.X., and X.S.Z. conducted the investigation, analyzed the results, and contributed to the visualization. Y.W. and A.F.A. drafted the manuscript. Y.W., A.F.A., W.L., W.X., and X.S.Z. revised and edited the manuscript. X.S.Z. secured the funding support and supervised the project.

\section*{COMPETING INTERESTS}

\paragraph{} The authors declare no competing interests.

\end{document}